\documentclass[twocolumn,aps,prb,showpacs]{revtex4}
\usepackage{graphicx}

\begin{document}

\title{Electronic structure and magnetic and optical properties of double
perovskite Bi$_2$FeCrO$_6$ from first-principles investigation}

\author{Zhe-Wen Song$^{1,2}$ and Bang-Gui Liu$^{1}$}
\affiliation{1. Beijing National Laboratory for Condensed Matter
Physics, Institute of Physics,  Chinese Academy of Sciences,
Beijing 100190, China\\
2. Department of Physics, Peking University, Beijing 100871,
China}

\date{\today}

\begin{abstract}
Double perovskite Bi$_2$FeCrO$_6$, related with BiFeO$_3$, is very
interesting because strong ferroelectricity and high magnetic
Curie temperature beyond room temperature are observed in it.
However, existing density-functional-theory (DFT) studies, using
pseudo-potentials, produce metallic ground state under the local
density approximation (LDA) and need LDA+U method to yield needed
nonmetallic ground state, resulting in low magnetic Curie
temperature (below 130 K). Here, we optimize its crystal structure
and then investigate its electronic structure and magnetic and
optical properties by combining the full-potential augmented plane
wave method with Monte Carlo simulation. Our optimized structure
is a robust ferrimagnetic semiconductor. This nonmetallic phase is
formed due to crystal field splitting and spin exchange splitting,
in contrast to Mott-Hubbard states in previous DFT studies. Spin
exchange constants and optical properties are calculated. Our {\it
ab initio} magnetic Curie temperature is 450 K, much higher than
previous DFT-based value and consistent with experimental results.
Our study and analysis reveals that the main magnetic mechanism is
an antiferromagnetic superexchange between Fe and Cr over the
intermediate O atom. These results are useful to understanding
such perovskite materials and exploring their potential
applications.
\end{abstract}


\maketitle

\section{Introduction}

Magnetic materials keeping high spin polarization at room
temperature or higher are highly desirable for spintronic
applications\cite{wolf,cro2-96}. Half-metallic materials\cite{hm}
and ferromagnetic (or ferrimagnetic) semiconductors are
intensively studied for this purpose. Up to 96\% spin polarization
has been achieved experimentally in the cases of (La,Sr)MnO$_3$
and CrO$_2$ materials\cite{cro2-96,lsmo}. Since 1998, double
perovskite oxides have been extensively explored. Both half-metals
and semiconductors with macroscopic magnetization beyond room
temperature have been found in such
materials\cite{oxide,la2fecro6,sum1,sum2,lsmo1}. High quality
samples of semiconductor technology standard have been achieved
for some of these compounds\cite{sr2crreo6}. Furthermore, it has
been established that ferroelectricity and magnetism can coexist
above room temperature in BiFeO$_3$ materials
\cite{bifeo3a,bifeo3b,bifeo3c}. These stimulate world-wide
interest in exploring multiferroic materials for novel
devices\cite{bifeo3a,bifeo3b,bifeo3c,tbmno3}.

Double perovskite Bi$_2$FeCrO$_6$ can be constructed from BiFeO$_3$,
and is first predicted to have both ferroelectricity and
ferrimagnetism, with theoretical electric polarization 80
$\mu$C/cm$^2$ and magnetic Curie temperature below 130
K\cite{bi2fecro6,bi2fecro6a}. Unfortunately, the Mott-Hubbard ground
state excludes higher Curie temperature in such theoretical studies.
Subsequently, the Bi$_2$FeCrO$_6$ is synthesized successfully by
several groups, and experimental measurements on good samples reveal
electric polarization up to 60 $\mu$C/cm$^2$ and magnetic Curie
temperature well above room temperature
\cite{exp1,exp2,exp3,exp4,exp5,exp6,exp7}. The best experimental
Curie temperatures are much higher than the first-principles
predicted one, while the electric polarization is consistent between
theory and experiment. Reasonable explanation of this contradiction
is highly desirable to understanding the basic physics of such
materials and exploring their applications.

In this paper, we combine a full-potential density-functional-theory
(DFT) method with Monte Carlo simulation to investigate double
perovskite Bi$_2$FeCrO$_6$, with emphasis on its electronic
structure and magnetic Curie temperature. We shall first optimize
its structure with GGA and then study its electronic structure. We
shall show that the resultant ferrimagnetic semiconductor phase is
robust against both change of exchange-correlation potential and
variation of structural parameters. Furthermore, we shall calculate
spin exchange constants and thereby determine the magnetic Curie
temperature. Our {\it ab initio} magnetic Curie temperature is 450
K, much higher than previous DFT-based value but consistent with
experimental results. In addition, we shall present calculated
optical properties. The magnetic mechanism is understood in terms of
the electronic properties. More detailed results will be presented
in the following.

The rest of this paper will be organized as follows. In the next
section, we shall describe our computational details. In the third
section we shall describe first-principles structural optimization
and present our optimized results. In the fourth section we shall
present main first-principles electronic structures. In the fifth
section we shall present first-principles magnetic calculations
and resulting Curie temperature. In the sixth section we shall
present first-principles optical properties. In the seventh
section we shall make some necessary discussions. Finally, we
shall give our conclusion in the eighth section.

\section{Computational details}

We use the full-potential augmented plane wave method within the
density functional theory (DFT) \cite{dft}, as implemented in
package WIEN2k\cite{wien2k}. Generalized gradient approximation
(GGA)\cite{pbe96} is used for the exchange-correlation potential
for calculating our main results, and local density approximation
(LDA)\cite{pw92} and a modified Becke-Johnson exchange potential
(mBJ)\cite{mbj} are also used for comparative calculations. Full
relativistic effects are calculated with the Dirac equations for
core states, and the scalar relativistic approximation is used for
the other states\cite{relsa,relsa1,relsa2}. The spin-orbit
coupling is neglected because it has little effect on our main
conclusion (to be detailed in the following). The muffin-tin radii
of Bi, Fe, Cr, and O are set to 2.13, 1.84, 1.84, and 1.64 bohr.
The spherical expansion in the muffin-tin spheres is done up to
$l_{max}$=10. The parameter $R_{mt}\times K_{max}$ is set to 7.5.
We use 1500 k-points in the first Brillouin zone (226 k-points in
the irreducible Brillouin zone) during the calculations. The
self-consistent calculations are considered to be converged only
when the integration of absolute charge-density difference per
formula unit between the successive loops is less than
$0.0001|e|$, where $e$ is the electron charge.

Metropolis algorithm and variants are used for our Monte Carlo
simulations\cite{mc,metropolis}. Several three-dimensional
lattices of up to  30$\times$30$\times$30 NaCl unit cells with
periodic boundary condition are used in these calculations. The
first 50,000 Monte Carlo steps (MCS) of total 100,000 MCS are used
for the thermal equilibrium, and the remaining 50,000 MCS are used
for a given temperature. The $T_c$ value is determined through
investigating the average magnetization and magnetic
susceptibility as functions of temperature\cite{mc}.

\section{Structural optimization}

It has already been proved\cite{bi2fecro6,bi2fecro6a} that the
lowest energy structure of double perovskite Bi$_2$FeCrO$_6$ has
R3 symmetry (space group \# 146). Its structural difference from
the cubic double perovskite is mainly due to the alternating
rotations of oxygen octahedra around Fe/Cr cations. The magnetic
order turns out to be ferrimagnetic with every Cr spin orienting
upward and every Fe spin downward. We use experimental lattice
constant $a=5.537$ \AA{} and $c=13.502$ \AA{}
($\alpha=60.20^{\circ}$) which are intermediate values of existing
experimental results\cite{exp1,exp2,exp3,exp4,exp5,exp6,exp7}.
After internal structure optimization, every atom has a
displacement from that of the cubic perovskite structure. We show
its crystal structure in Fig. 1. The optimized parameters are
summarized in Table \ref{table1}. The bond lengthes and Fe-O-Cr
bond angle can be calculated from the structural parameters. The
calculated values are summarized in Table \ref{table2}.

\begin{figure}[!htb]
\includegraphics[width=4cm]{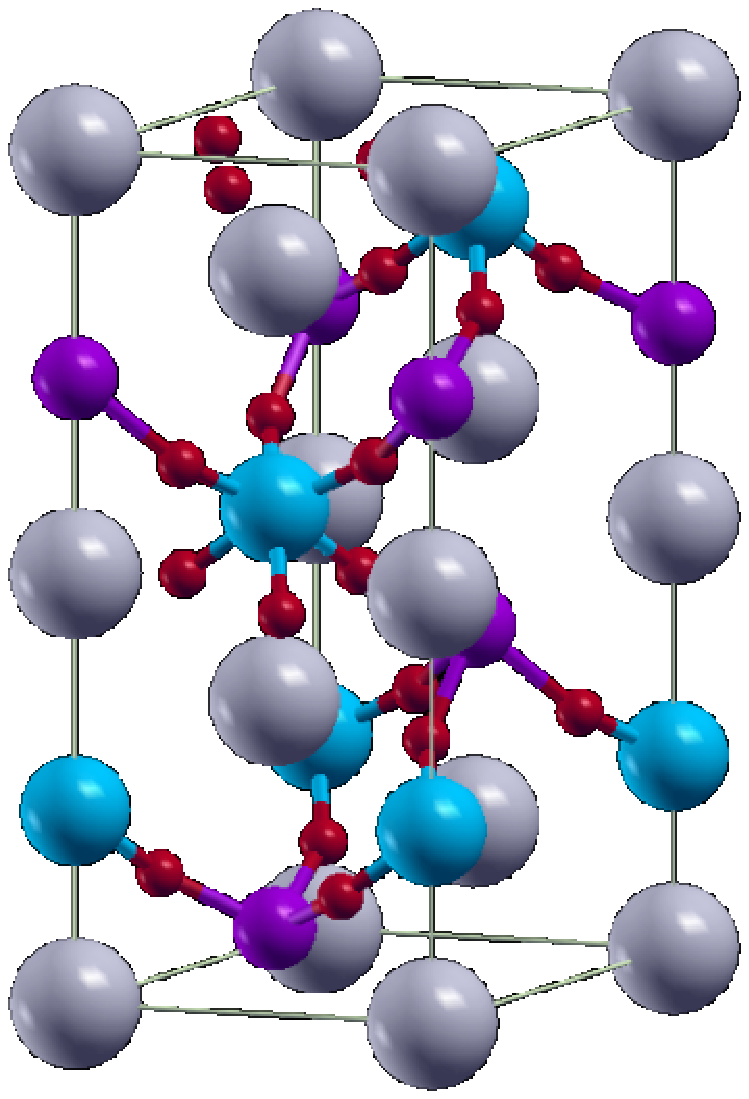}
\includegraphics[width=4cm]{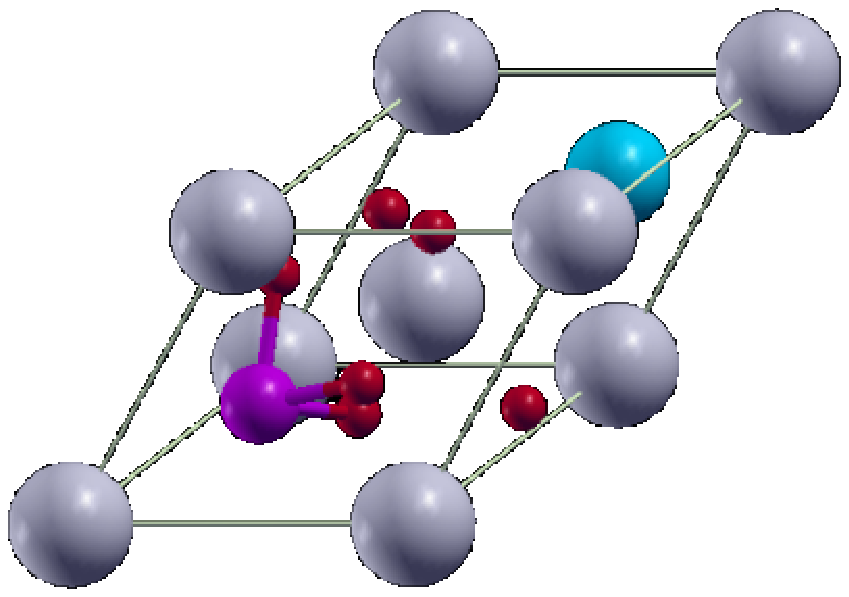}
\caption{(color online) The Rc (\#146) Crystal structure of double
perovskite Bi$_2$FeCrO$_6$. The right-side is the primitive cell,
and the left-side is used to describe the relative positions of
the atoms. O, Fe, Cr, and Bi are denoted by colorful balls with
different size from the smallest to the biggest. }\label{struct}
\end{figure}

\begin{table}[htb]
\caption{Optimized internal structural parameters of double
perovskite Bi$_2$FeCrO$_6$ with the Rc (\#146) crystal
structure.}\label{table1}
\begin{ruledtabular}
\begin{tabular}{ccccc}
& Atom & Parameter & Values  & \\
\hline
& Bi    & $z_1$ & 0.0005, 0.5040 & \\
\hline
& Fe/Cr & $z_2$ & 0.7330/0.2260 & \\
\hline
& O     &  $x$  & 0.5433, 0.0511 & \\
&       &  $y$  & 0.9542, 0.9043 & \\
&       &  $z$  & 0.3937, 0.4459 &
 \end{tabular}
 \end{ruledtabular}
\end{table}

These structural parameters are not much different from the
previous LDA-optimized results\cite{bi2fecro6,bi2fecro6a}.
Actually, we obtain 55.2$\mu$C/cm$^2$ for the sum $\sum_i z_iu_i$
of formal charge $z_i$ times displacement $u_i$, using formal
charges 3.0, 3.0, 3.0, and -2.0 for Bi, Fe, Cr, and O,
respectively. This value is almost the same as the previous
55$\mu$C/cm$^2$ in Refs. 15 and 16. Therefore, we believe that
similar ferroelectric polarization can be calculated with the
present structural parameters. The total magnetic moment is
equivalent to 2$\mu_B$ per formula unit. Different magnetic
moments, 3.70$\mu_B$ and 2.18$\mu_B$, are observed in the spheres
of Fe and Cr atoms, and magnetic moments in the spheres of other
atoms are tiny. These implies that Fe and Cr are both in the high
spin states.

\begin{table}[htb]
\caption{Bond lengths (\AA) and Fe-O-Cr angle ($^{\circ}$) of
double perovskite Bi$_2$FeCrO$_6$ with the Rc (\#146) crystal
structure.}\label{table2}
\begin{ruledtabular}
\begin{tabular}{ccccc}
& Bond length      & Fe-O  & 2.130, 1.971 & \\
&       & Cr-O  & 1.947, 1.994 & \\
&       & Bi-O  & 2.347, 2.387 & \\
&       &       & 2.406, 2.416 & \\
\hline
&  Bond angle     &  Fe-O-Cr  & 153.7, 153.6 & \\
 \end{tabular}
 \end{ruledtabular}
\end{table}

\section{Electronic Structure}

With the optimized crystal structure, we can study the electronic
structure of the Bi$_2$FeCrO$_6$. In Fig. 2 we present the
spin-resolved total density of states (DOS) and partial DOS
projected in the d states of Fe and Cr between -7.3 and 5 eV. The
corresponding spin-dependent band structure is presented in Fig.
3. It is clear that the Bi$_2$FeCrO$_6$ is a ferrimagnetic
semiconductor, in contrast with the ferrimagnetic metallic phase
obtained in previous pseudo-potential
calculations\cite{bi2fecro6,bi2fecro6a,dft3,dft4}.

\begin{figure}[!htb]
\includegraphics[width=7cm]{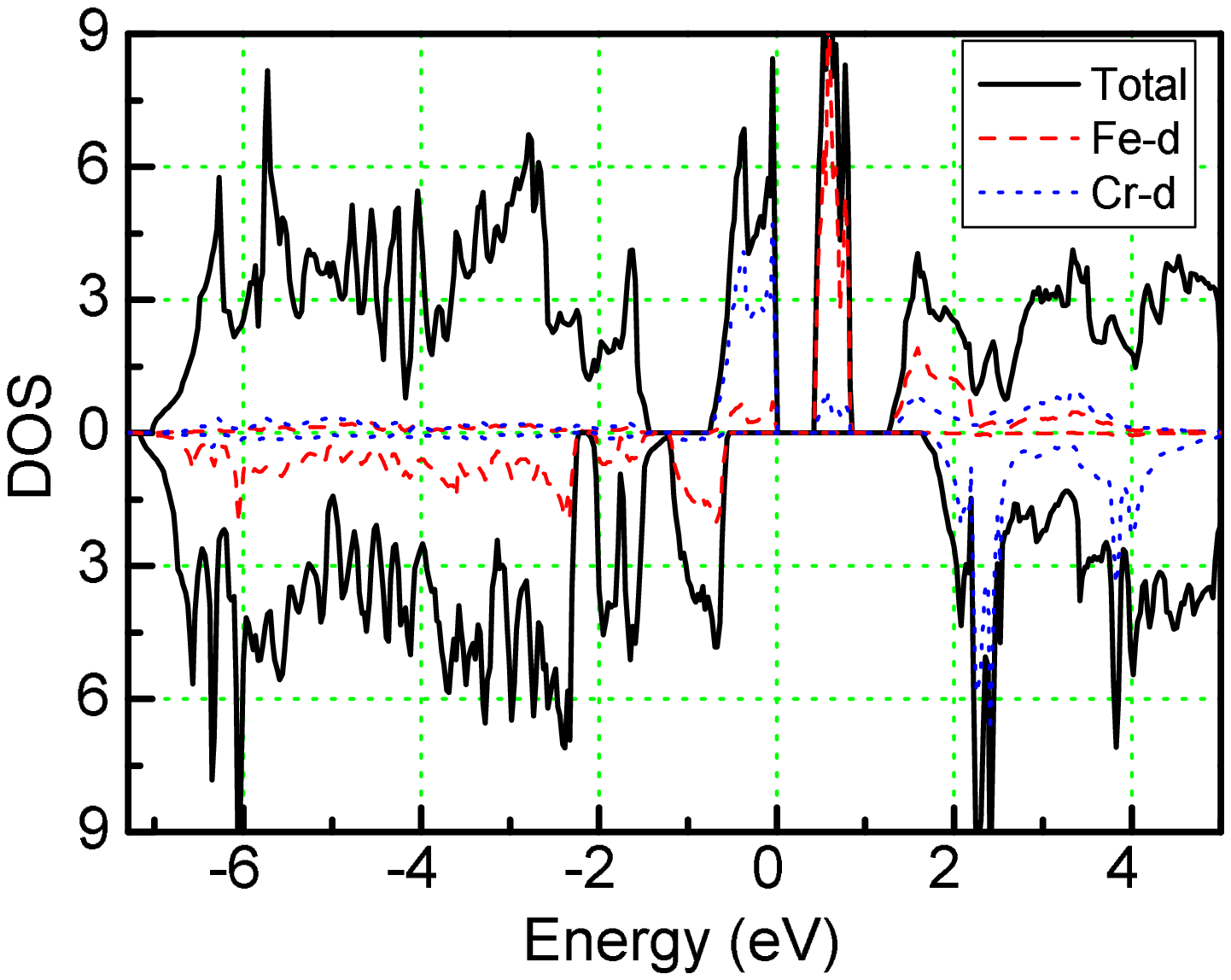}
\caption{(color online) Spin-resolved density of states (DOS, in
state/eV per formula unit) of double perovskite Bi$_2$FeCrO$_6$,
calculated with GGA. The solid line is total DOS, and dashed and
dotted lines refer to partial DOS projected in the atomic spheres
of Fe and Cr, respectively. The upper part in each panel is
majority-spin DOS result, and the lower the minority-spin
one.}\label{dos_gga}
\end{figure}

\begin{figure}[!htb]
\includegraphics[height=5cm]{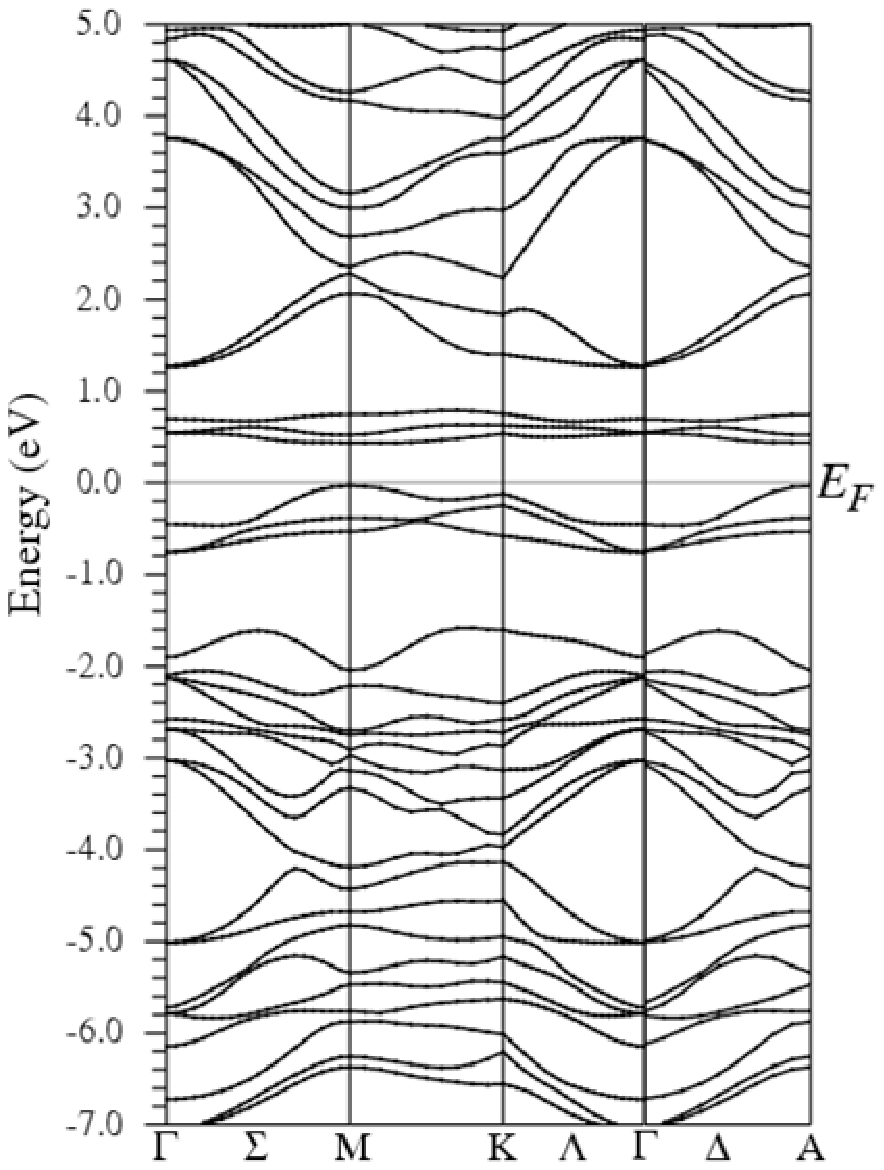}
\includegraphics[height=5cm]{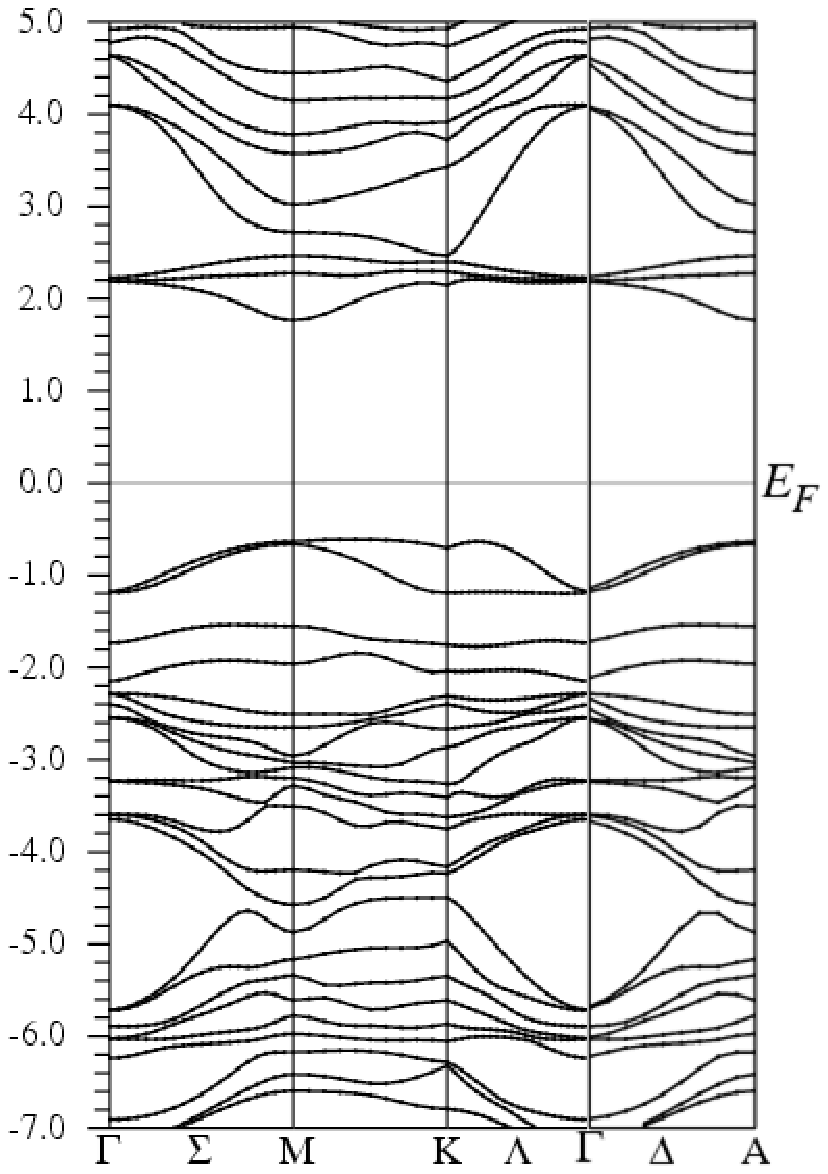}
\caption{(color online) Spin-resolved energy bands of double
perovskite Bi$_2$FeCrO$_6$, calculated with GGA. The left-side
panel is for majority-spin, and the right-side for
minority-spin.}\label{band_gga}
\end{figure}

In the majority-spin channel, the bands between -7.3 and -1.4 eV
are from the 18 O 2p states; the three bands between -0.8 and 0 eV
originate mainly from Cr 3d t2g; the three empty bands are almost
Fe 3d t2g states; and the upper states are from Fe 3d eg, Cr 3d
eg, and Bi 6p. In the minority-spin channel, the 23 filled bands
are from 18 O 2p states and 5 Fe 3d ones; and the empty bands from
Cr 3d and Bi 6p states. Here, the Fe 3d states hybridize with the
O 2p and the Cr 3d ones interact strongly with the Bi 6p, in
contrast to the majority-spin channel.

Because it is well known that both LDA and GGA underestimate
semiconductor gap, we use a new exchange potential, mBJ, to
improve the electronic structure, especially the semiconductor
gap. The mBJ DOS are presented in Fig. 4. The main features are
similar to the GGA results, except that the mBJ gap is much wider
than the LDA and GGA ones. The three semiconductor gaps are
summarized in Table \ref{table3}.

\begin{figure}[!htb]
\includegraphics[width=7cm]{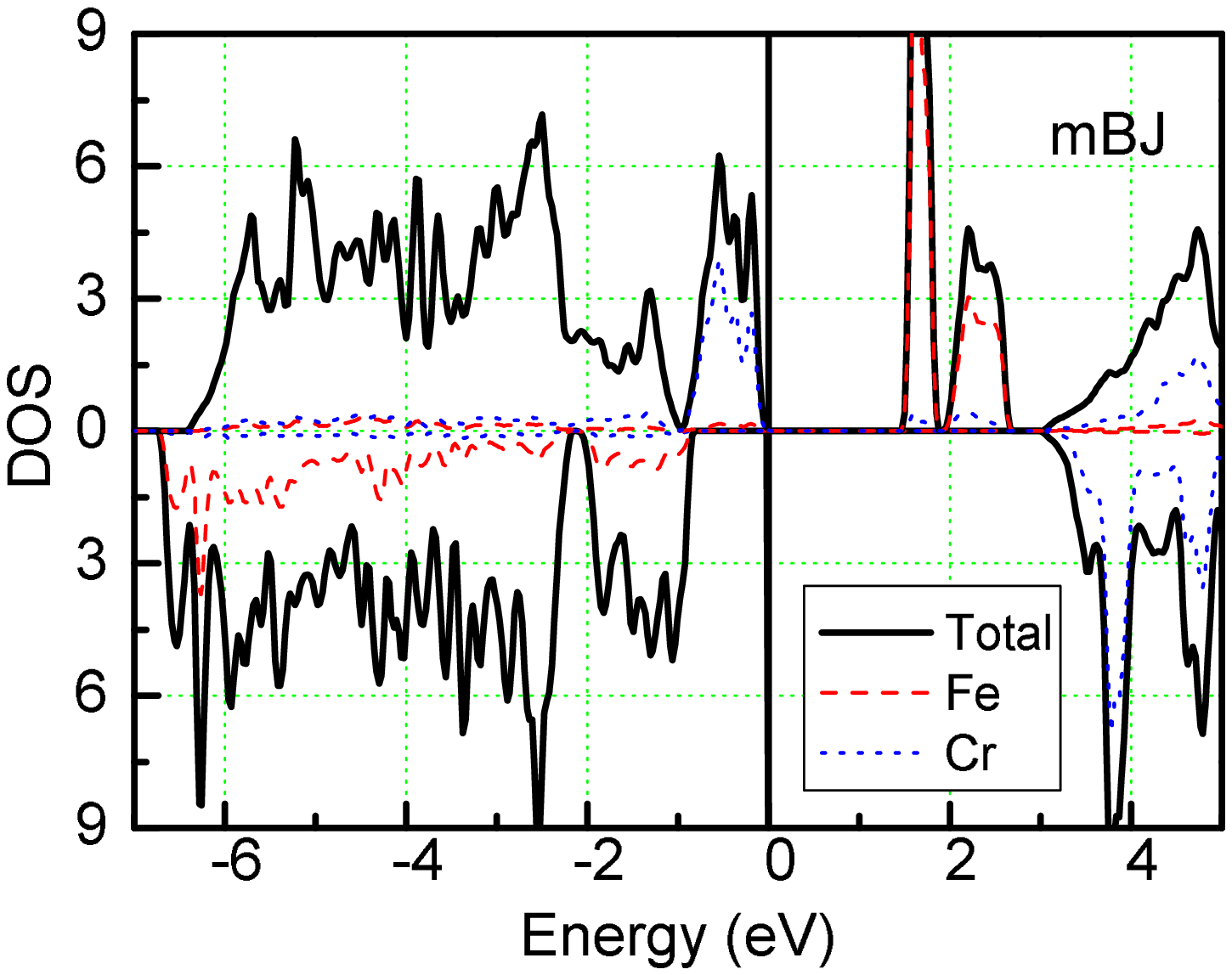}
\caption{(color online) Spin-resolved density of states (DOS, in
state/eV per formula unit) of double perovskite Bi$_2$FeCrO$_6$,
calculated with mBJ. The solid line is total DOS, and dashed and
dotted lines refer to partial DOS projected in the atomic spheres
of Fe and Cr, respectively. The upper part in each panel is
majority-spin DOS result, and the lower the minority-spin
one.}\label{dos_lda}
\end{figure}

\begin{table}[htb]
\caption{The semiconductor energy gaps (in eV) of double
perovskite Bi$_2$FeCrO$_6$, calculated with our optimized
structure and previous one\cite{bi2fecro6,bi2fecro6a} in terms of
LDA, GGA, and mBJ.}\label{table3}
\begin{ruledtabular}
\begin{tabular}{cccccc}
& Structure & LDA  & GGA  &  mBJ  & \\
\hline
& this &  0.26  &  0.42 &    1.48   & \\
\hline
& previous\cite{bi2fecro6,bi2fecro6a} &  0.30   & 0.48  &  & \\
 \end{tabular}
 \end{ruledtabular}
\end{table}

Our study shows that this feature of ferrimagnetic semiconductor
is robust to a changing of crystal structure parameters. For
example, the GGA and LDA gaps in the lower line of Table
\ref{table3} are calculated with the previous structure
parameters\cite{bi2fecro6,bi2fecro6a}.

\section{Magnetic Curie temperature}

In order to study spin exchange interactions, we compare the total
energies of the four magnetic configurations in the supercell
consisting of 20 atoms, namely ferrimagnetic (the ground state)
and ferromagnetic orders, and two other magnetic orders
constructed by reversing one of the two Cr or Fe spins in the
ferrimagnetic order. With respect to the ground state, the three
magnetic structures have total energies: 310, 161, and 166 meV per
formula unit, respectively. Because the induced spin density at
the other atoms are very small compared to those at the magnetic
atoms, we consider only Fe and Cr. The magnetic energies reflect
the spin exchange energies $e_{ij}$ where $i$ and $j$ denote two
different spins. Fe and Cr atoms form a lattice of the NaCl
crystal structure\cite{oxide,sum1,sum2,exp1,exp4}. While the
magnetic moments in the spheres of Fe and Cr are 3.70 and 2.18
$\mu_B$, the Fe$^{3+}$ and Cr$^{3+}$ cations contribute 5 and 3
$\mu_B$, respectively. We can assign spin values $s=\frac 52$ and
$s=\frac 32$ to the Fe and Cr spins, respectively. Therefore, we
obtain the effective spin Hamiltonian:
\begin{equation}
H=\sum_{\langle ij\rangle}J_{ij}\vec{S}_i\cdot \vec{S}_j
\end{equation}
where $\vec{S}_i$ is spin operator at site $i$ (in both of the Fe
and Cr sublattices), the summation is over spin pairs, and the
spin exchange constant $J_{ij}$ is limited to the nearest (Fe-Cr)
and the next nearest (Fe-Fe and Cr-Cr) spin pairs. There exists a
relation between the energies $e_{ij}$ and constants $J_{ij}$,
$e_{ij}=J_{ij}s_is_j$, where $s_i$ takes either $\frac 52$ or
$\frac 32$. The spin exchange energies $e_{ij}$ and constants
$J_{ij}$ are summarized in Table \ref{table4}.

\begin{table}[htb]
\caption{The calculated spin exchange energies ($e_{ij}$) and
constants ($J_{ij}$) of double perovskite Bi$_2$FeCrO$_6$ for the
nearest and next nearest spin pairs ($\langle ij\rangle$), and the
resulting Curie temperature $T_c$ from Monte Carlo
simulation.}\label{table4}
\begin{ruledtabular}
\begin{tabular}{cccccc}
& $\langle ij\rangle$ & Fe-Cr  & Fe-Fe  &  Cr-Cr  & \\
\hline
& $e_{ij}$ (meV) &  25.8  &  -1.88 &    -1.06   & \\
\hline
& $J_{ij}$ (meV) &  6.89  &  -0.30 &    -0.47   & \\
\hline \hline
& $T_c$ (K) &  450  &   &    & \\
 \end{tabular}
 \end{ruledtabular}
\end{table}


\begin{figure}[!htb]
\includegraphics[width=7cm]{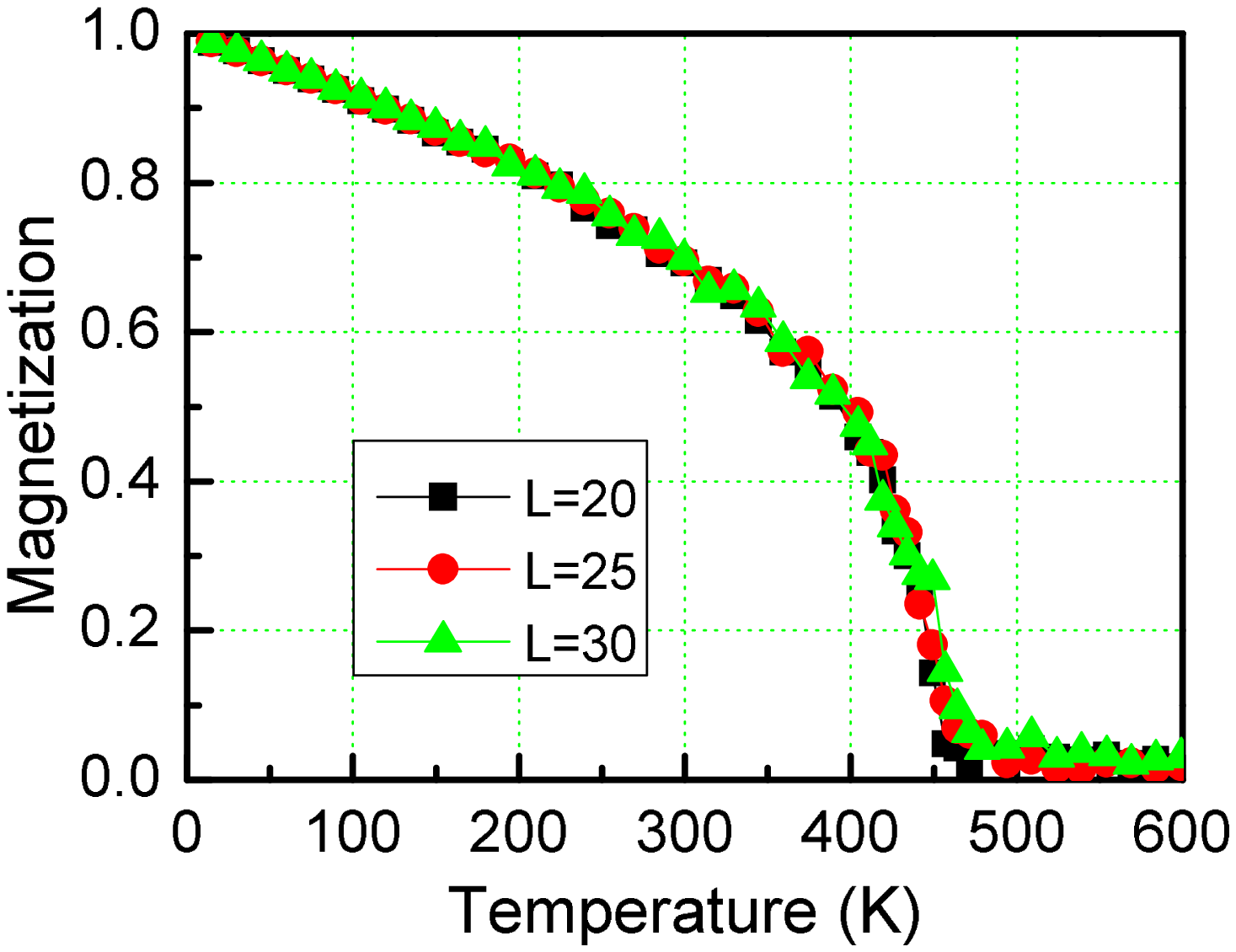}
\caption{(color online) Temperature dependent magnetization (in
unit of 2$\mu_B$ per Fe-Cr pair) of double perovskite
Bi$_2$FeCrO$_6$, simulated with Monte Carlo method and the
classical Heisenberg model.}\label{mag}
\end{figure}

We carry out Monte Carlo simulations of the Heisenberg model to
estimate the $T_c$ of the materials\cite{mc,metropolis}. It is
well known that Curie temperature will be a little underestimated
if classical approximation to the Heisenberg model is used in the
Monte Carlo simulation. We use Monte Carlo simulation of the
classical Heisenberg model to give a lower bound for the Curie
temperature of the quantum spin model (1). Our simulated average
magnetization (in unit of 2$\mu_B$ per Fe-Cr pair) as a function
of temperature from classical Heisenberg model is presented in
Fig. 5. Because the classical approximation is used, the
low-temperature part is not very accurate with respect to the
model (1), but the high-temperature part should be accurate enough
to get reasonable Curie temperature. The calculated $T_c$ value is
equivalent to 450 K, meaning that real Curie temperature should be
a little higher than 450 K. This Curie temperature is much higher
than the original prediction (lower than 130
K)\cite{bi2fecro6,bi2fecro6a}, but it is consistent with
experimental high Curie temperatures beyond room
temperature\cite{exp1,exp4,exp5,exp6,exp7}.

\section{Optical properties}

Because optical properties are important to semiconductor, we
calculate the direct electronic contributions of the optical
dielectric function $\epsilon(\omega)$ and optical conductivity
$\sigma(\omega)$ curves of the Bi$_2$FeCrO$_6$ as functions of
photon energy ($\omega$). Considering that the semiconductor gaps
have important impact on the low-energy parts of these optical
functions, we calculate them with both GGA and mBJ. These results,
both real (Re) and imaginary parts (Im) between 0 and 12 (or 13)
eV, are presented in Figs. 6 and 7.

\begin{figure}[!htb]
\includegraphics[width=7cm]{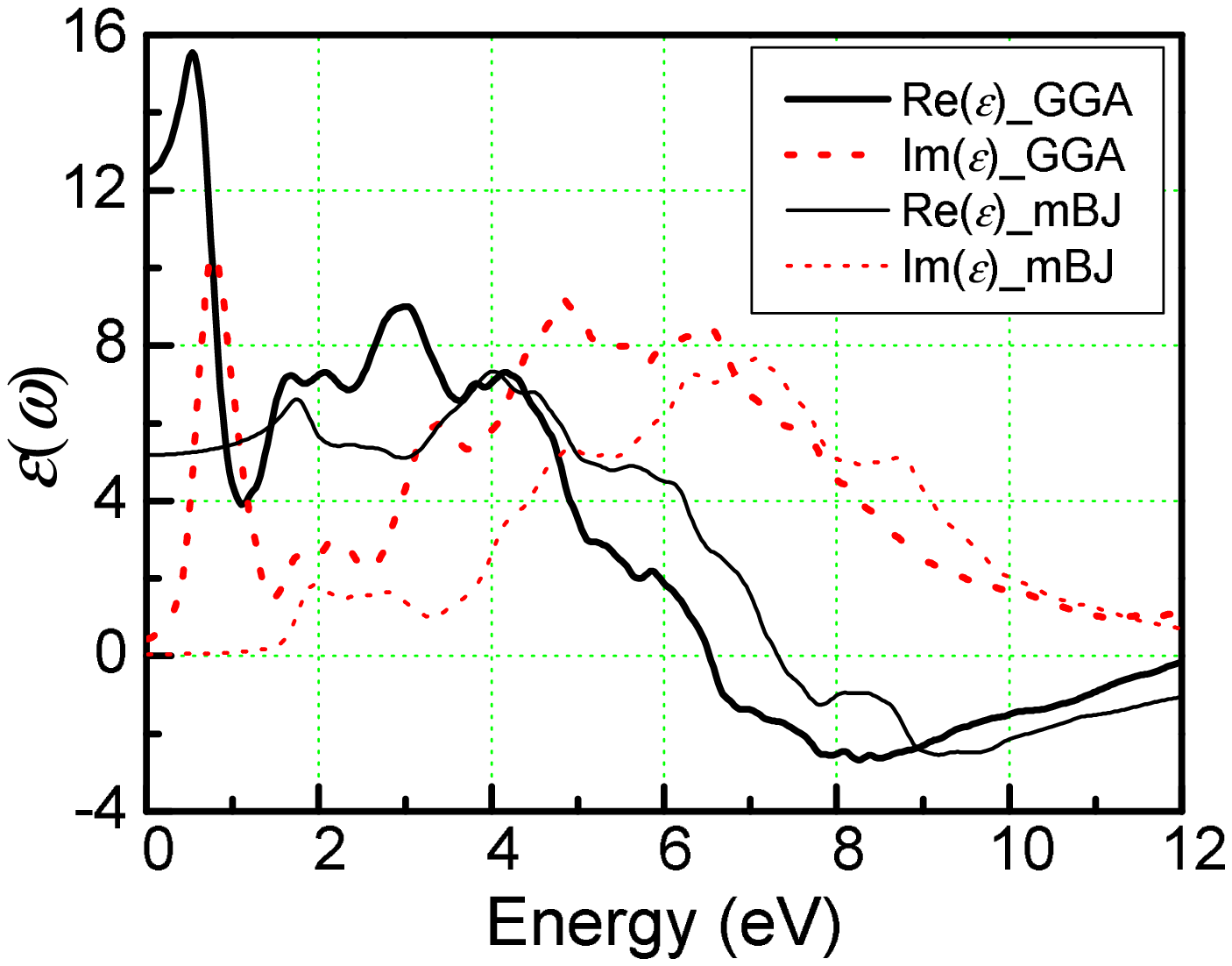}
\caption{(color online) Optical dielectric functions, both real
(solid) and imaginary (dotted) parts, of double perovskite
Bi$_2$FeCrO$_6$, calculated with GGA (thick) and mBJ
(thin).}\label{epsilon}
\end{figure}
\begin{figure}[!htb]
\includegraphics[width=7cm]{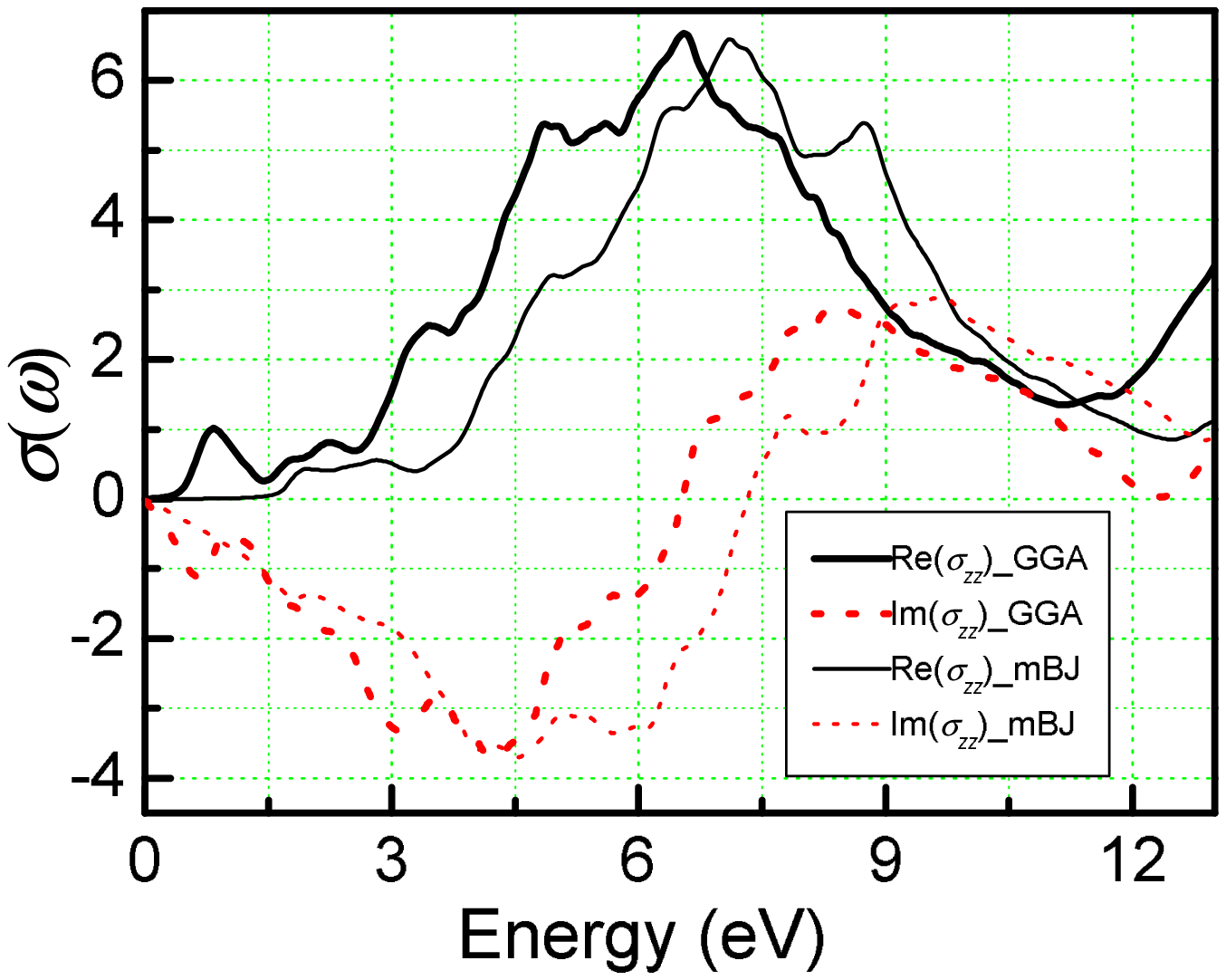}
\caption{(color online) Optical electric conductivity, both real
(solid) and imaginary (dotted) parts, of double perovskite
Bi$_2$FeCrO$_6$, calculated with GGA (thick) and mBJ
(thin).}\label{sigma}
\end{figure}

For Re($\epsilon$), the GGA result of the low energy limit is
substantially larger than the mBJ one, both have similar trend
when the photon energy is larger than 1.5 eV. For Im($\epsilon$),
main difference appears for low energy (0-1.5 eV). There is a
sharp peak at 0.75 eV for the GGA result, which is related to the
steep decrease in the GGA curve of Re($\epsilon$) at the same
photon energy. The mBJ result of Im($\epsilon$) is equivalent to
zero in this region, reflecting the broad semiconductor gap from
mBJ. For higher energy, both of the results are similar. For the
$\sigma(\omega)$ curves, the main difference also appears in the
low energy region, Re($\sigma(\omega)$) having a peak
approximately at 1.0 eV and Im($\sigma(\omega)$) having a valley
approximately at 0.5 eV. For the higher energy, both of the
$\sigma(\omega)$ curves have similar behaviors. These calculated
optical functions could be useful in exploring the optical
properties of the Bi$_2$FeCrO$_6$ materials and others similar.

\section{Discussions}

Our DFT investigation with GGA, LDA, and mBJ shows that the
ground-state phase of double perovskite Bi$_2$FeCrO$_6$ is a
ferrimagnetic semiconductor with a clear semiconductor gap (larger
than 0.26 eV), in contrast to previous ferrimagnetic metallic
phases from LDA pseudo-potential
calculations\cite{bi2fecro6,bi2fecro6a,dft3,dft4}. Even using the
previous structural paramters\cite{bi2fecro6,bi2fecro6a}, we still
obtain the ferrimagnetic semiconductor phase with both GGA and
LDA. Furthermore, our spin exchange energies and interaction
constants produce the high Curie temperature 450 K which is
consistent with experimental high Curie temperature beyond room
temperature\cite{exp1,exp4,exp5,exp6,exp7}. This is in contrast to
the previous LDA+U calculations which yield Mott-Hubbard insulator
phases and low Curie temperatures (below 130
K)\cite{bi2fecro6,bi2fecro6a}. These imply that we need not use
LDA+U or GGA+U to achieve the reasonable ferrimagnetic
semiconductor phase for double perovskite Bi$_2$FeCrO$_6$.

\begin{figure}[htbp]
\includegraphics[height=5cm]{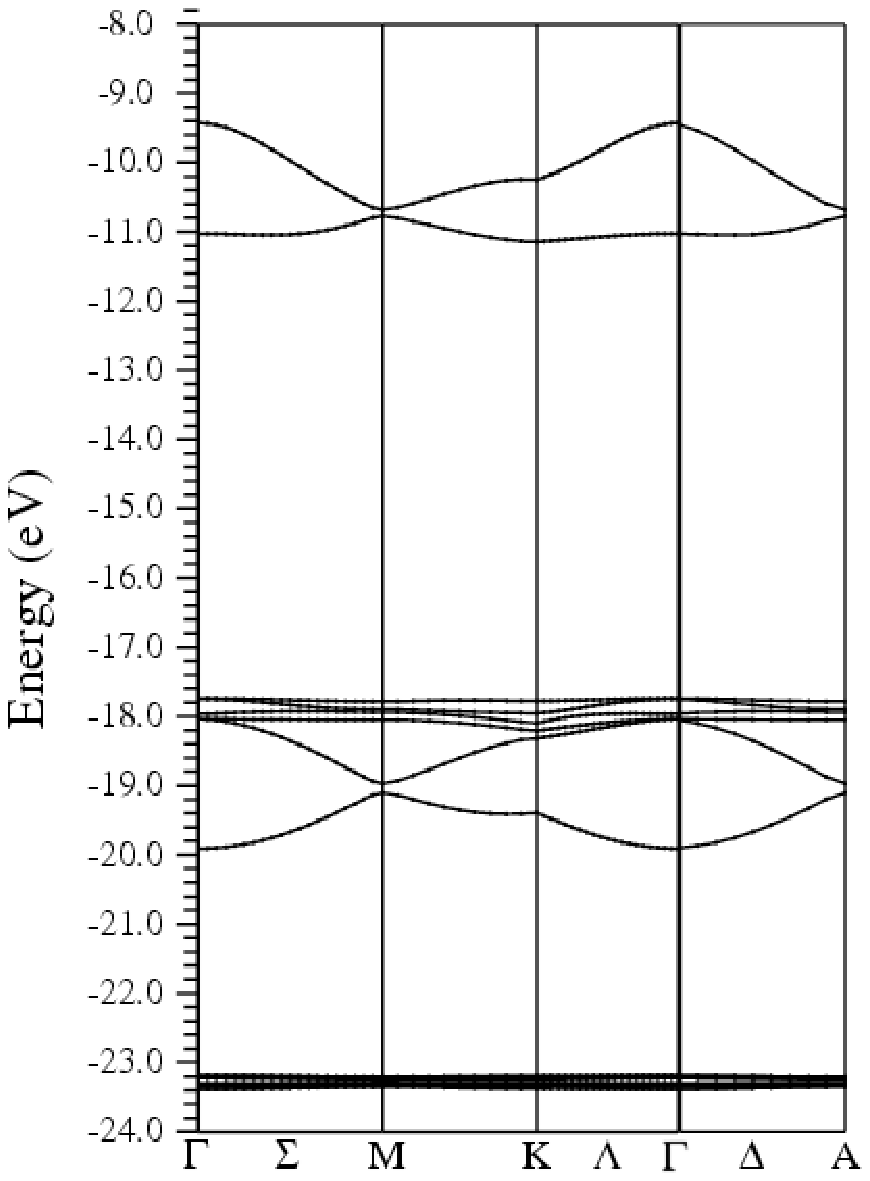}
\includegraphics[height=5cm]{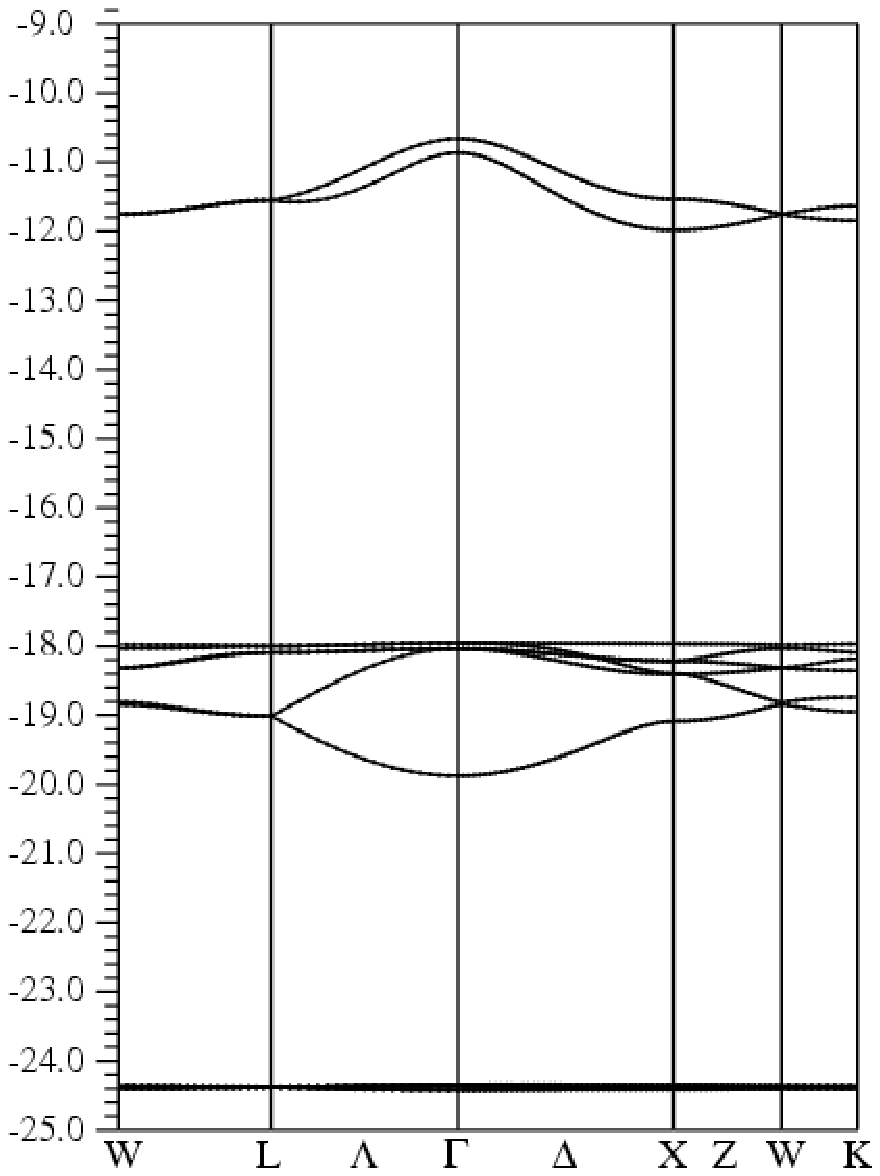}
\caption{(color online) The majority-spin part of the Bi 6s, O 2s,
and Bi 5d energy bands of double perovskite Bi$_2$FeCrO$_6$ in the
R3 (the left part) and cubic (the right part) structures,
calculated with GGA. }\label{band2}
\end{figure}

Here, our semiconductor gap is formed between the filled Cr 3d t2g
and the empty Fe 3d t2g bands. Essentially, it is originated
mainly from the crystal field splitting due to the deformation of
the O octahedrons plus the spin exchange splitting of the 3d
electrons. Our calculated spin exchange interaction between the
nearest Fe and Cr atoms is positive, and the next nearest (Fe-Fe
and Cr-Cr) ones are negative. These spin interactions become
substantially weaker from the nearest to next nearest neighboring
spin pairs, which implies that such a spin description is
reliable. The main spin interaction is intermediated by the O atom
in between the nearest Fe and Cr atoms. The Fe and Cr spins
contribute opposite magnetic moments so that the ferrimagnetism is
formed. Therefore, the main magnetic mechanism is the
antiferromagnetic superexchange over the O atom.

Now we address possible contributions from deeper energy levels.
For this purpose, we present in Fig. 8 the bands of Bi 6s, O 2s,
and Bi 3d states of the R3 (space group \#146) and cubic (space
group \#225) structures. They are below -9 eV from the Fermi
levels. The Bi 5d bands (-23.3 or -24.4 eV) in both of the
structures are completely flat, meaning that Bi 5d states are
isolated, but there is a clear correlation between the two Bi 6s
bands (from -11.0 to -9.4 eV) and two of the six O 2s ones (from
-19.9 to -17.7 eV) in the case of the R3 structure. Noticeably,
its Bi 6s bands are more separated from the O 2s ones than those
of the cubic structure. This can be attributed to the expelling
due to the small but nonzero hybridization of the two sets of s
states. The interaction between the Bi 6s and O 2s states should
be involved in the ferroelectric property in the R3 structure,
because they are absent in the cubic structure without
ferroelectricity.

\section{Conclusion}

In summary, we have investigated the electronic structure and
magnetic and optical properties of double perovskite
Bi$_2$FeCrO$_6$ by combining the full-potential augmented plane
wave method with Monte Carlo simulation. Our calculations show
that our full-potential GGA-optimized phase is a ferrimagnetic
semiconductor and it is robust against both change of
exchange-correlation potential and variation of structural
parameters. This nonmetallic phase is formed due to crystal field
splitting and spin exchange splitting, in contrast to Mott-Hubbard
insulating state in previous DFT studies. Spin exchange constants
and optical properties are calculated. Our {\it ab initio}
magnetic Curie temperature is 450 K, much higher than previous
DFT-based value but consistent with experimental results. Our
study and analysis reveals that the main magnetic mechanism is an
antiferromagnetic superexchange between Fe and Cr over the
intermediate O atom. Our results are useful to understanding such
materials and exploring their potential applications.

\begin{acknowledgments}
This work is supported by Nature Science Foundation of China
(Grant Nos. 11174359 and 10874232) and by Chinese Department of
Science and Technology (Grant No. 2012CB932302).
\end{acknowledgments}

\end{document}